\documentclass[a4paper,fleqn,usenatbib]{mnras}

\usepackage{mathptmx}
\usepackage{amssymb}
\usepackage[T1]{fontenc}
\usepackage{ae,aecompl}
\usepackage{adjustbox}

\usepackage{graphicx}	
\usepackage{amsmath}	
\usepackage{amssymb}	
\usepackage{natbib} 

\title[Polarimetry through multi-line observations]{Chromospheric polarimetry through multi-line observations of the 850 nm spectral region}

\author[C. Quintero Noda et al.]{C. Quintero Noda,$^{1}$\thanks{E-mail: carlos@solar.isas.jaxa.jp}
T. Shimizu,$^{1}$
Y. Katsukawa,$^{2}$
 J. de la Cruz Rodr\'{i}guez,$^{3}$
\newauthor
M. Carlsson,$^{4}$
T. Anan,$^{5}$
T. Oba,$^{6}$
K. Ichimoto,$^{2,5}$
Y. Suematsu$^{2}$ 
\\
$^{1}$Institute of Space and Astronautical Science, Japan Aerospace Exploration Agency, Sagamihara, Kanagawa 252-5210, Japan\\
$^{2}$National Astronomical Observatory of Japan, 2-21-1 Osawa, Mitaka, Tokyo 181-8588, Japan\\
$^{3}$Institute for Solar Physics, Dept. of Astronomy, Stockholm University, Albanova University Center, SE-10691 Stockholm, Sweden\\
$^{4}$Institute of Theoretical Astrophysics, University of Oslo, P.O. Box 1029 Blindern, N-0315 Oslo, Norway\\
$^{5}$Kwasan and Hida Observatories, Kyoto University, Kurabashira Kamitakara-cho, Takayama-city, 506-1314 Gifu, Japan\\
$^{6}$SOKENDAI, Shonan Village, Hayama, Kanagawa 240-0193 Japan\\
}

\date{Accepted XXX. Received YYY; in original form ZZZ}

\pubyear{2016}

\begin{document}
\label{firstpage}
\pagerange{\pageref{firstpage}--\pageref{lastpage}}
\maketitle

\begin{abstract}
Future solar missions and ground-based telescopes aim to understand the magnetism of the solar chromosphere. We performed a supporting study in \cite{QuinteroNoda2016} focused on the infrared Ca~{\sc ii} 8542 \AA \ line and we concluded that is one of the best candidates because it is sensitive to a large range of atmospheric heights, from the photosphere to the middle chromosphere. However, we believe that it is worth to try improving the results produced by this line observing additional spectral lines. In that regard, we examined the neighbour solar spectrum looking for spectral lines that could increase the sensitivity to the atmospheric parameters. Interestingly, we discovered several photospheric lines that greatly improve the photospheric sensitivity to the magnetic field vector. Moreover, they are located close to a second chromospheric line that also belongs to the Ca~{\sc ii} infrared triplet, i.e. the Ca~{\sc ii} 8498 \AA \ line, and enhances the sensitivity to the atmospheric parameters at chromospheric layers. We conclude that the lines in the vicinity of the  Ca~{\sc ii} 8542 \AA\ line not only increase its sensitivity to the atmospheric parameters at all layers, but also they constitute an excellent spectral window for chromospheric polarimetry.
\end{abstract}




\section{Introduction}

Space missions have traditionally focused on performing spectropolarimetric observations measuring the four Stokes parameters in a narrow spectral window where one or two photospheric absorption lines of interest are present. For instance, see Hinode/SP \citep{Tsuneta2008,Lites2013}, SDO/HMI  \citep{Pesnell2012,Schou2012}, and Solar Orbiter/PHI  \citep{Gandorfer2011,Solanki2015}. On the contrary, ground-based telescopes usually have instruments that can cover several spectral lines simultaneously as THEMIS/MTR \citep{LopezAriste2000,Paletou2001}, SST/CRISP \citep{Scharmer2003,Scharmer2008}, DST/IBIS \citep{Cavallini2006} and DST/SPINOR \citep{SocasNavarro2006}, Gregor/GRIS \citep{Schmidt2012,Collados2012}, or ZIMPOL \citep{Povel2001}, among others. In most of the mentioned cases, the light beam occupies almost the full length of the camera in one of its directions due to the larger spectral coverage what directly increases the amount of data generated.  Although the data rate is not usually a crucial factor for ground-based telescopes, space-based missions have an extremely limited telemetry. Thus, unless there is a strong reason to expand the wavelength coverage it is highly recommendable to keep focusing on narrow spectral windows that contain a high density of useful lines. In this regard, we find it extremely helpful to perform theoretical studies and observations to define the optimum spectral window for the purposes of a given mission. For example, Hinode/SP was strongly supported by ASP \citep{Elmore1992} observations and SDO/HMI benefited from ground-based observations but also from specific supporting works as \cite{Norton2006}. We performed a similar study in \cite{QuinteroNoda2016} aiming to support future missions with chromospheric polarimetry as the main target, for instance, Solar-C \citep{Katsukawa2011,Katsukawa2012,Watanabe2014,Suematsu2016}. We concluded that the Ca~{\sc ii} 8542~\AA \ is a unique spectral line that is sensitive to a large range of heights, from the photosphere to the chromosphere (see Figure 4 and 6 of the mentioned work). However, its sensitivity to the magnetic field at photospheric layers is low  which precludes examining quiet Sun magnetic fields at these heights. Therefore, there is still room for improvement and, for this reason, we decided to examine the solar spectrum at the vicinity of the Ca~{\sc ii} 8542~\AA \ line. In this regard, if we observe additional spectral lines we will increase the number of spectral points with valuable information, particularly if these lines have different heights of formation, which will enhance the accuracy of the inferred atmospheric parameters \citep[for example, ][]{AsensioRamos2007}.  However, in order to achieve this purpose we have to expand the spectral window which will increase the data rate if we maintain the same spectral sampling. Therefore, it is essential that these additional lines provide valuable information complementing the Ca~{\sc ii} 8542~\AA \ spectral line.

The aim of this work is to provide a thorough study of one of the most promising spectral windows to perform simultaneous polarimetric observations of the photosphere and chromosphere. This is of high relevance for future missions as Solar-C, but also to future ground-based telescopes as DKIST \citep{Elmore2014} or EST \citep{Collados2013}.

\section{Spectral lines}

\subsection{Characterization}\label{spectral}

We have thoroughly examined the spectral lines that can be detected from 840 to 870 nm using a solar atlas \citep{Delbouille1973} looking for promising candidates. We found the three chromospheric lines from the Ca~{\sc ii} infrared triplet located at 8498, 8542, and 8662 \AA, and a large amount of photospheric lines. Moreover, we also wanted to find a spectrum that could be observed with one single camera, without changing the wavelength range. Therefore, we have to bear in mind that the instrument's camera has available a finite number of pixels. In that sense, the wavelength spectrum we can fit in it is proportional to the spectral sampling ($\Delta\lambda$) we use, i.e. the larger the spectral sampling (poorer spectral resolution) the larger the spectrum we can observe. Thus, we realized that we cannot observe a spectral range of at least 170~\AA \ (required to cover the Ca~{\sc ii} infrared triplet)  with a conventional camera of 2000 or 4000 pixels without largely increasing the spectral sampling. As the Ca~{\sc ii} 8542~\AA \ line is the best of the triplet (see, for instance, \cite{Pietarila2007,delaCruzRodriguez2012}), we looked for interesting lines in its vicinity. We discovered that there are more promising lines at bluer wavelengths for two main reasons. The first one is that, contrary to the Ca~{\sc ii} 8662~\AA \ line, the Ca~{\sc ii} 8498~\AA \ line core is not blended with any other solar line, which allows to unambiguously separate the line core chromospheric contribution from any type of parasitic contribution\footnote{There is an iron line that is blended with the Ca~{\sc ii} 8498~\AA \ line but is located at -1.03~\AA \ from its line core, while in the case of the Ca~{\sc ii} 8662~\AA \ line there is an iron line that is located at -0.24~\AA \ from its line centre.}. The second reason is that the region around the Ca~{\sc ii} 8498~\AA \ line presents two strong absorption lines whose effective Land\'{e} factor is higher than 1.75. 

 \begin{figure}
\begin{center} 
 \includegraphics[trim=20 30 -5 0,width=8.7cm]{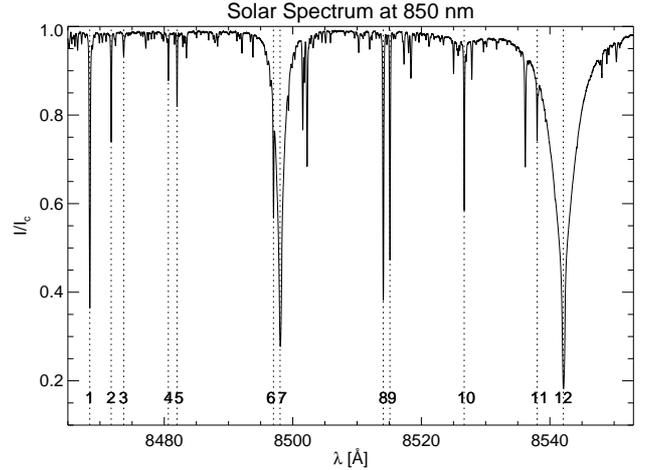}
 \caption{Solar spectrum for the infrared region located around 850~nm. We highlighted the most important lines with vertical dotted lines and we labelled them with numbers. These numbers correspond to the ones used in Table \ref{table_lines}.}
 \label{atlas}
 \end{center}
\end{figure}

\begin{table} 
   \hspace{-0.5cm}
  \begin{adjustbox}{width=0.49\textwidth}
  \begin{tabular}{cccccccc}
	\hline
 & 	Atom                   & $\lambda$ [\AA] & $\log$ gf & $L_l$       & $U_l$           & $g_{eff}$ & $I_{core}$ [au] \\
	\hline
1      & Fe~{\sc i}             &  8468.41      & -2.072 & ${}^5P_{1}$      &  ${}^5P^{0}_{1}$      & 2.50 & 3649    \\
2      & Fe~{\sc i}             &  8471.74      & -0.915 & ${}^5D^{0}_{3}$  &  ${}^5D_{3}$          & 1.50  & 7510    \\
3      & Mg~{\sc i}             &  8473.69      & -2.020 & ${}^3P^{0}_{2}$  &  ${}^3S_{1}$          & 1.25 & 9347    \\	
4      & Fe~{\sc i}             &  8480.63      & -1.685 & ${}^5D^{0}_{2}$  &  ${}^5P_{1}$          & 1.00  & 8784    \\
5      & Fe~{\sc i}             &  8481.98      & -1.647 & ${}^3F_{2}$      &  ${}^3D^{0}_{1}$      & 0.75 & 8230    \\
6      & Fe~{\sc i}             &  8496.99      & -0.950 & ${}^3F^{0}_{3}$  &  ${}^3F_{2}$          & 1.50  & 5831    \\
7      & Ca~{\sc ii}            &  8498.02      & -1.318 & ${}^2D_{3/2}$    &  ${}^2P_{3/2}$        & 1.07 & 2856    \\
8      & Fe~{\sc i}             &  8514.07      & -2.229 & ${}^5P_{2}$      &  ${}^5P^{0}_{2}$      & 1.83 & 3840    \\
9      & Fe~{\sc i}             &  8515.11      & -2.073 & ${}^3G_{3}$      &  ${}^3G^{0}_{3}$      & 0.75 & 4734    \\
10     & Fe~{\sc i}             &  8526.67      & -0.760 & ${}^5D^{0}_{4}$  &  ${}^5D_{4}$          & 1.50  & 5858    \\
11     & Fe~{\sc i}             &  8538.01      & -1.400 & ${}^5D^{0}_{4}$  &  ${}^7G_{4}$          & 1.40  & 7423    \\
12     & Ca~{\sc ii}            &  8542.09      & -0.363 & ${}^2D_{5/2}$    &  ${}^2P^{0}_{3/2}$    & 1.10 & 1823    \\
	\hline
  \end{tabular}   
  \end{adjustbox}
\caption{Spectral lines included in the infrared 850 nm window. Each column, from left to right contains the number we assigned to each line in Figure \ref{atlas}, the corresponding atomic species, line core wavelength centre, the oscillator strength of the transition, the spectroscopic notation of the lower and the upper level, the effective Land\'{e} factor, and the line core intensity in arbitrary units (the continuum level corresponds to 10000).}\label{table_lines} 
\end{table}

We can see the spectral lines corresponding to the mentioned 850 nm window in Figure \ref{atlas}. We labelled with numbers the lines we selected to perform this study and we describe them in detail in Table \ref{table_lines}. Their atomic information was retrieved from the NIST \citep{nist2015} and VALD \citep{vald2015} databases, and \cite{SocasNavarro2007}. These lines fulfil the Russel-Saunders (or L-S) coupling scheme and we computed their effective Land\'{e} factor as

\begin{equation}
g_{eff}=\frac{1}{2}(g_{u}+g_{l})+\frac{1}{4}(g_{u}-g_{l})[J_{u}(J_{u}+1)-J_{l}(J_{l}+1)],
\end{equation}
where $g_{u}$ and $J_{u}$ are the Land\'{e} factor and the total angular momentum of the upper level of the transition while $g_{l}$ and $J_{l}$ correspond to the same quantities of the lower level of the transition. The Land\'{e} factor of each level is computed as 

\begin{equation}
g=\frac{3}{2}+\frac{S(S+1)-L(L+1)}{2J(J+1)},
\end{equation}
where $S$, $L$, and $J$, are the total spin, orbital angular momentum and total angular momentum of the corresponding level. For more information, see \cite{Landi2004}.

Besides the stronger lines, we also take into account weaker spectral lines that are located close to the line core of the Ca~{\sc ii} infrared lines, see lines number 6 and 11, in order to check if these lines can contaminate the Ca~{\sc ii} lines. The rest of the unlabelled lines in Figure \ref{atlas} are weak, they cannot be described under the L-S coupling scheme, or they are not clearly defined in the mentioned databases.

If we examine Figure \ref{atlas}, we will quickly notice the two wide Ca~{\sc ii} lines, labels 7 and 12, but also two deep narrow lines that posses a large effective Land\'{e} factor, see numbers 1 and 8 in Table \ref{table_lines}. In fact, these numbers are relatively close to those of the commonly used iron pair located at 630 nm, i.e. Fe~{\sc i} 6301.5 ($g_{eff}=1.67$) and Fe~{\sc i} 6302.5 ($g_{eff}=2.5$). Moreover, as the Zeeman splitting is proportional to the square of the wavelength (see Eq. 3.14 of \cite{Landi2004}) these two lines are theoretically (the amplitude of the Stokes polarization parameters also depends on the gradient of the source function)  more proficient for measuring the Stokes polarization profiles produced by weak magnetic fields.

\subsection{Methodology}

We have found several promising line candidates based on their effective Land\'{e} factor. However, we want to quantify their capabilities and their sensitivity to the magnetic field using different atmospheric models as input. We aim to synthesize the four Stokes parameters assuming non-local thermodynamic equilibrium (NLTE) conditions using the {\sc nicole} code \citep{SocasNavarro2015} that optimally works for the infrared Ca~{\sc ii}. The code has the limitation that only  one atomic species can be treated in NLTE at the same time. However, the only two NLTE lines that appear in Table \ref{table_lines} belong to transitions of the same atom, therefore, we can synthesize all the lines of the mentioned table simultaneously, computing the rest of the lines under the LTE approximation. We employed {\sc nicole} for computing the Stokes profiles using a semi-empirical 1D model (see Section \ref{1dmodel}) and a realistic 3D model (see Section \ref{3dmodel}).

\begin{figure}
\begin{center} 
 \includegraphics[trim=10 3 7 0,width=7.5cm]{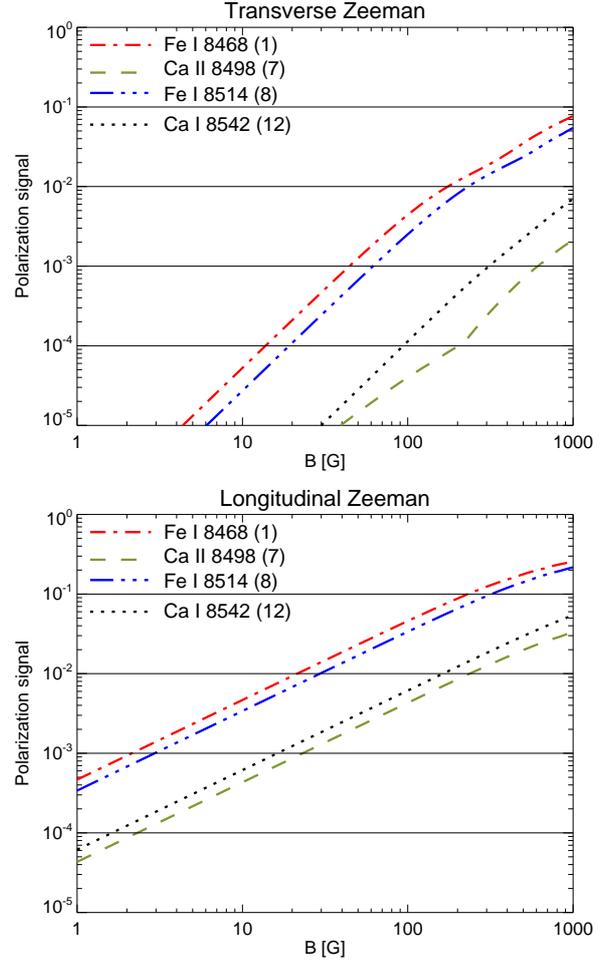}
 \caption{Maximum polarization signals for the two strongest photospheric lines (see number 1 and 8 in Table \ref{table_lines}) and the two Ca~{\sc ii} lines.  Upper panel shows the total linear polarization while the lower panel displays the maximum Stokes $V$ signals.}
 \label{Pol_signals}
 \end{center}
\end{figure}

\section{Synthesis of Stokes profiles using a 1D model}\label{1dmodel}

\subsection{Maximum polarization signals}

The first study we performed was computing the polarization signals produced by the lines of Table \ref{table_lines} using a semi-empirical model. We followed the same procedure used in \cite{QuinteroNoda2016}  (see section 5.3 of that work) where we added a constant magnetic field to the FALC atmospheric model \citep{Fontenla1993} and we changed its strength from 1 to 1000 G. For simplicity, we used a single inclination and azimuth value of 45 degrees and we only modified the magnetic field strength. 

\begin{figure*}
\begin{center} 
 \includegraphics[trim=10 0 -20 0,width=17.5cm]{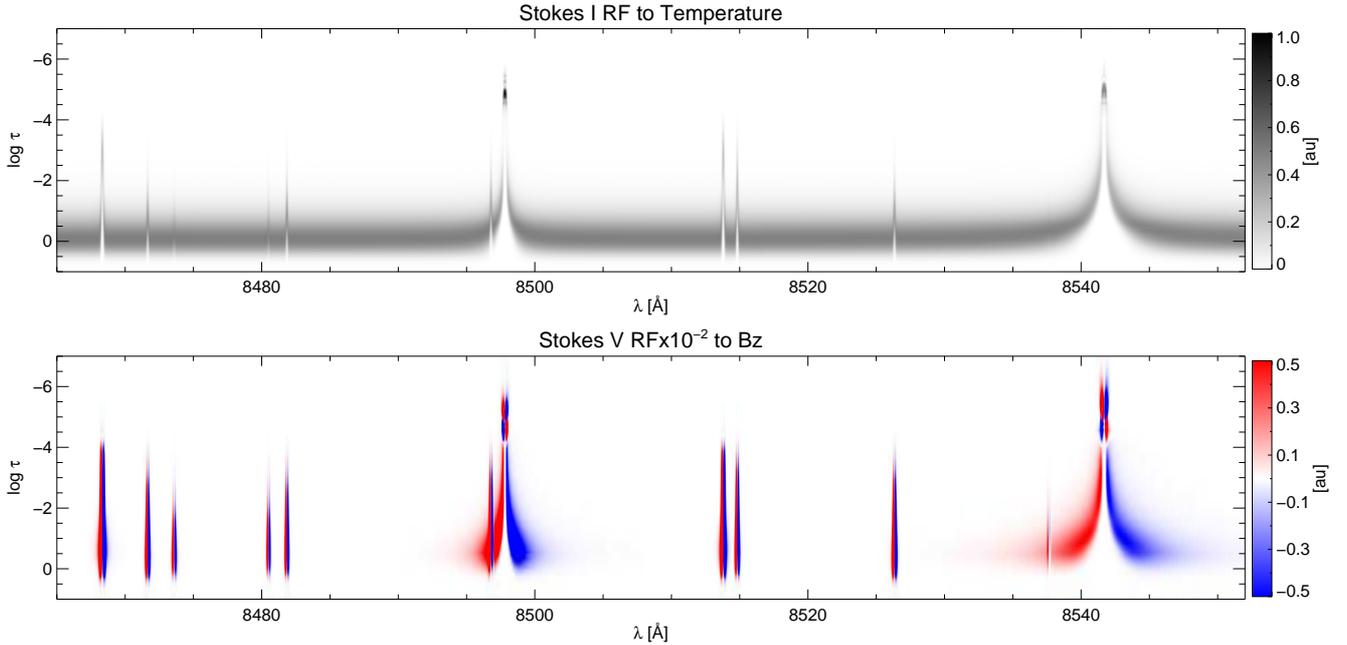}
 \vspace{-0.2cm} 
 \caption{Two-dimensional plot of the Stokes $I$ RF to changes in the temperature (top) and Stokes $V$ RF to perturbations in the longitudinal field (bottom). White areas indicate regions of no sensitivity to changes in the atmospheric parameters, while red and blue colours indicate opposite signs of the RF. The length of the spectrum is 88 \AA \ and it covers all the lines of Table \ref{table_lines}. Both panels are normalized to the maximum of the Stokes $I$ RF to temperature.}
 \label{2DRF}
 \end{center}
\end{figure*}

The results for the most relevant lines\footnote{Although Table \ref{table_lines} contains 12 lines we opted to focus only on the most important ones, i.e. the two chromospheric Ca~{\sc ii} lines and the strongest photospheric iron lines (labels 1 and 8),  to facilitate the visualization of the different studies we performed.} are included in Figure \ref{Pol_signals}. We can see that the polarization signals are different for the two Ca~{\sc ii} lines (black and green colours), being always larger for the  Ca~{\sc ii} 8542 \AA \ line, especially for the transverse field (top panel). These differences could come from the slightly larger Land\'{e} factor of the latter but also to different Zeeman splitting patterns as it is also a much broader line (see Figure \ref{atlas}), although we should perform additional studies in the future to confirm these hypothesis. These results reveal that we can detect inclined fields with a strength as low as $\sim120$ G with a noise level of $5\times10^{-4}$ of the continuum intensity signal ($I_c$). On the other hand, if we focus on the photospheric lines (red and blue colours) we can see that the one with largest $g_{eff}$, i.e. Fe~{\sc i} 8468 \AA, always shows larger polarization signals. In addition, for both cases we can detect inclined fields whose strength is as low as $20$~G with the same reference noise level.

\subsection{Response functions}

The next study we did is to determine the sensitivity of these lines to changes in the atmospheric parameters following the same procedure used in \cite{QuinteroNoda2016} (see Section 3) to compute the response functions (RF) for the lines included in Table \ref{table_lines}. We opted in this case to reduce the number of 2D plots because the spectral coverage is too large to properly see the RF if we use the same 20 panels included in Figure 6 of the mentioned paper. Thus, we just show in Figure~\ref{2DRF} the Stokes~$I$ RF to the temperature and the Stokes $V$ RF to the longitudinal magnetic field, both normalized to the maximum value of the first response function.

\begin{figure*}
\begin{center} 
 \includegraphics[trim=-5 6 0 0,width=17.8cm]{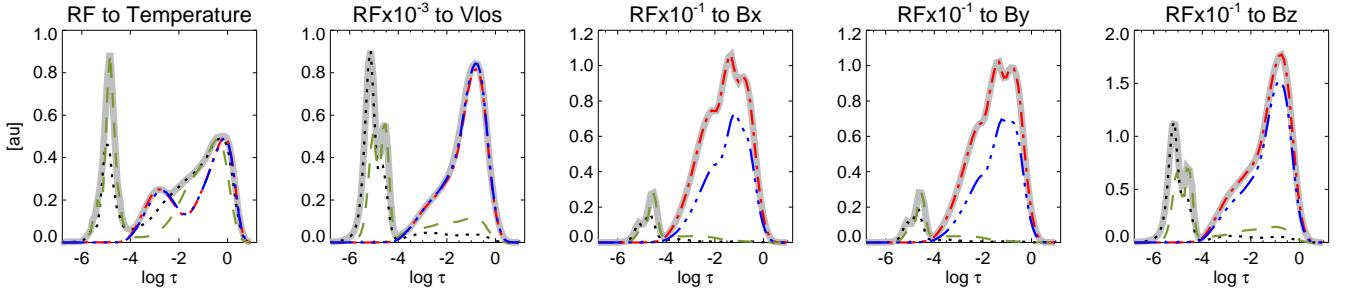}
 \vspace{-0.2cm}
 \caption{Maximum RF, from left to right, for the temperature, LOS velocity, and the three components of the magnetic field vector. Each curve corresponds to the RF maximum value for the four Stokes parameters and for all the computed wavelengths for the four strongest lines in Table \ref{table_lines}, i.e. Fe~{\sc i} 8468 \AA, Ca~{\sc ii} 8498 \AA, Fe~{\sc i} 8514 \AA, and Ca~{\sc ii} 8542 \AA. We used the same colour and line type criteria used in Figure \ref{Pol_signals}. The effective sensitivity from combining the results for all the lines of Table \ref{table_lines} is plotted in grey.}
 \label{1DRF}
 \end{center}
\end{figure*}

Starting with the temperature RF, top panel, we can see that the two Ca~{\sc ii} lines reach higher in the atmosphere, up to approximately $\log \tau=-5.5$ while the photospheric lines, mainly the Fe~{\sc i} 8468 and 8514 \AA \ lines, show sensitivity up to $\log \tau=-3.8$. Similar results can be found for the longitudinal field, bottom panel, where the asymmetric shape of the Stokes $V$ profile is clearly identified through opposite RF values for each spectral line wing (blue and red colours). In addition, we can also detect two white narrow bands around $\log \tau=[-4, -5]$ where the sensitivity to the longitudinal field drops to almost zero before the RF changes of sign. 

Additionally, it is also useful to study the vertical stratification of the RF at certain wavelengths as was shown, for instance, by \cite{RuizCobo1992} or \cite{Fossum2005}. For this reason, we made a one-dimensional plot computing the maximum of the RF to different physical parameters at any wavelength of the spectrum shown in Figure \ref{2DRF}. Moreover, we aim to compare these results with the ones obtained synthesizing only the Ca~{\sc ii} 8542 \AA \ line. Thus, we plotted the different possible scenarios in Figure \ref{1DRF}, where dotted colour lines designate the RF values for the four strongest lines of Table \ref{table_lines} while the solid grey line depicts the effective RF using the results of combining all the spectral lines.

If we start with the RF to perturbations in the temperature we find that there is an improvement at chromospheric layers thanks to the Ca~{\sc ii} 8498 \AA \ line (green) whose peak is larger although it is located at slightly lower heights. Unfortunately, there is not much improvement at lower layers as the Ca~{\sc ii} 8542 \AA \ wings are already very sensitive to the temperature. However, the situation changes for the rest of the atmospheric parameters where observing all the lines of Table \ref{table_lines} (grey) largely enhances the sensitivity to the atmospheric parameters with respect to the case of observing only the  Ca~{\sc ii} 8542 \AA \ line (black). This is partially because photospheric lines, i.e. red and blue colours, are highly sensitive to changes in the atmospheric parameters below $\log \tau\sim-3.8$, this improvement being even larger in the case of the horizontal components of the magnetic field (see third and fourth panels). In addition, we also detect a larger sensitivity to all the atmospheric parameters around $\log \tau\sim-4.5$ due to the Ca~{\sc ii} 8498 \AA \ line (green). Therefore, measuring all the lines presented in Table \ref{table_lines}, we do not only increase the sensitivity at photospheric layers but also in the chromosphere. Unfortunately, in spite of these good results, we still have a region of low sensitivity at $\log \tau\sim-4$ (see grey solid line) that now is clearer to detect due to the large sensitivity of photospheric lines. This low sensitivity region indicates that we will not be able to accurately determine the physical information of this atmospheric layer.

Finally, before moving to the next section, we would like to remark that the RF are model dependent. Therefore, we should consider the results of Figure \ref{2DRF} and \ref{1DRF} as illustrative findings instead of absolute ones.

\begin{figure*}
\begin{center} 
 \includegraphics[trim=15 0 15 0,width=17cm]{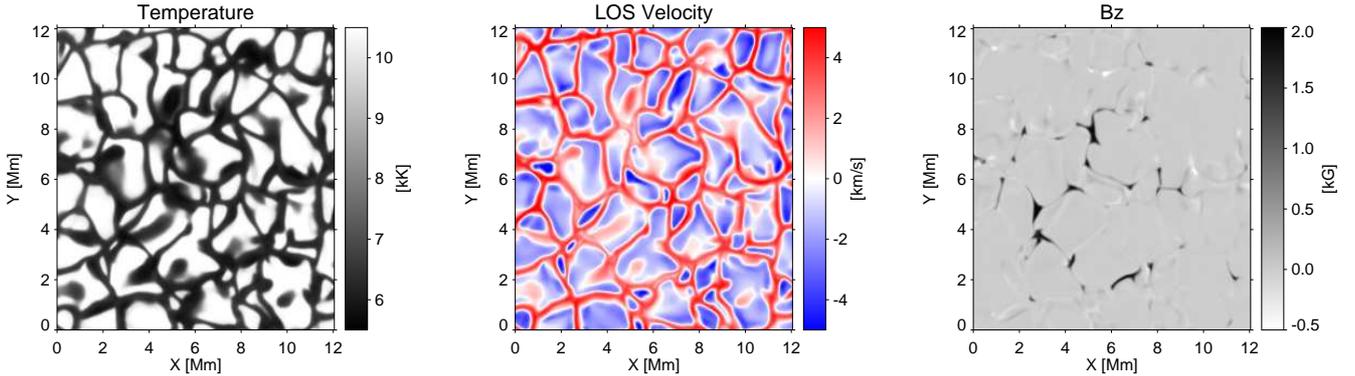}
 \vspace{-0.2cm}
 \caption{Simulation snapshot used in this work. It corresponds to a small fragment of the {\sc bifrost} snapshot 385. We show the spatial distribution of the atmospheric parameters at geometrical height Z$\sim$0 km. From left to right, temperature, LOS velocity, and longitudinal magnetic field.}
 \label{fov}
 \end{center}
\end{figure*}

\begin{figure*}
\centering
 \includegraphics[trim=4 4 0 0,width=17.8cm]{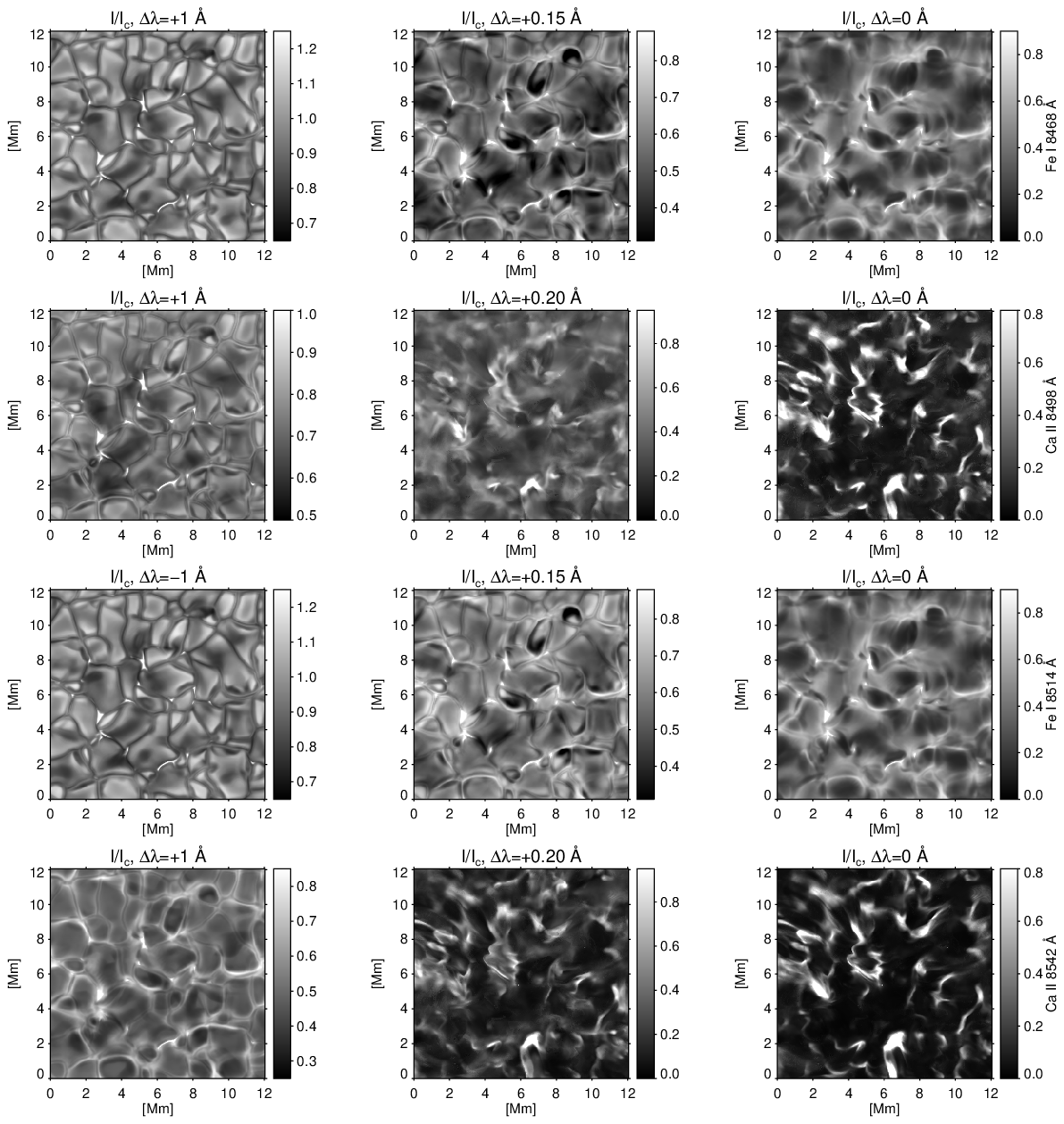}
 \vspace{-0.2cm}
 \caption{Spatial distribution of Stokes $I$ signals at different spectral positions along the lines of Table \ref{table_lines} with labels (from top to bottom) 1, 7, 8, and 12. Each column contains a different wavelength position from the line centre of the corresponding line. }
 \label{Intensity}
\end{figure*}

\section{Synthesis of Stokes profiles using a realistic 3D model}\label{3dmodel}

We employed in this section a snapshot from a 3D MHD simulation calculated with the {\sc bifrost} code \citep{Gudiksen2011} that includes a patch of enhanced network \citep{Carlsson2016} corresponding to snapshot number 385. This snapshot has been used in several studies focused on the Ca~{\sc ii} 8542 \AA \ \citep{delaCruzRodriguez2012,delaCruzRodriguez2013} and the Hanle effect produced by this line \citep{Stepan2016}, in the formation of the H$_{\alpha}$ line \citep{Leenaarts2012} and the nature of the H$_{\alpha}$ fibrils \citep{Leenaarts2015} or for studying the differences between the complete and partial redistribution assumption for the Mg~{\sc ii} h\&k lines \citep{Sukhorukov2016}.

We have selected a small fragment of the simulated snapshot, see Figure \ref{fov}. This fragment is different than the one used in \cite{QuinteroNoda2016}, i.e. it is three times larger and contains most of the magnetic patch located on the right side of the original field of view (see Figure 2 of \cite{Carlsson2016}). This magnetic field is located inside intergranular lanes that harbour large downflows and contain cool plasma. We chose a small fragment of the simulation in order to accelerate the process of synthesis but it is large enough to contain several granules and to allow us to study the spatial distribution of signals for different intergranular magnetic field concentrations.

We employed this simulation to examine the spatial distribution of intensity signals for different wavelengths and we compared these intensity signals for the most relevant lines of Table \ref{table_lines}. Later, we computed for those relevant lines their maximum polarization signals aiming to estimate their amplitude as their spatial distribution inside the network patches. The method employed is the same used in \cite{QuinteroNoda2016} where we used an ideal spectral sampling of 1 m\AA, a microturbulence value constant with height and equal to 3 km/s (see \cite{delaCruzRodriguez2012}), and we spectrally degraded the synthetic profiles using a macroturbulence of 1.5 km/s, i.e. we convolved the synthetic profiles with a Gaussian with 42 m\AA \ width, aiming to simulate realistic conditions.

\subsection{Line intensity at different wavelengths}

We studied the spatial distribution of intensity signals at different wavelength points along the spectral line for the four lines examined in the previous sections, i.e. labels 1, 7, 8, and 12 of Table \ref{table_lines}. We know that intensity images taken at the core of the Ca~{\sc ii} 8542~\AA \ line show fibrilar structures as the ones commonly found in H$_{\alpha}$ images although less prominent, \citep[e.g.,][]{delaCruzRodriguez2011,Rouppe2013}. In this study, we want to compare the landscape of intensity signals produced by the Ca~{\sc ii} 8498 \AA \ line with the one generated by the Ca~{\sc ii} 8542 \AA \ that, based on previous results, seems to form a bit higher in the chromosphere. Moreover, we expect that the rest of the relevant lines used in this study, i.e. the photospheric Fe~{\sc i} lines, will form lower in the atmosphere. However, we also want to examine the spatial distribution of their intensity signals to check how high in the photosphere these lines can probe the atmospheric parameters. 

We show in Figure \ref{Intensity} the intensity signals at three different wavelength positions from the line core of each selected line, i.e. $\Delta\lambda=[1,0.2,0]$~\AA \ from the line core for the Ca~{\sc ii} lines and  $\Delta\lambda=[1,0.15,0]$~\AA \ for the iron lines. In the case of the Fe~{\sc i} 8514~\AA, we selected $\Delta\lambda$=-1~\AA \ instead because there is a different spectral line at $\Delta\lambda$=+1~\AA \ from its line core (see Table \ref{table_lines}). If we focus first on the photospheric lines, i.e. odd rows, we can see the granulation pattern at $\Delta\lambda$=1~\AA \ (at $\Delta\lambda$=-1~\AA \ in the case of the Fe~{\sc i} 8514~\AA) from the line core (leftmost column), where granules are brighter than intergranular lanes. We also find bright intergranular lanes that correspond to regions where the longitudinal field is strong (see Figure \ref{3dmodel}) and produces a depression of the visible surface. Moving to a wavelength position closer to the line core, middle column, we start to detect the reverse granulation pattern \citep{Evans1972} where granules appear darker than intergranular lanes \citep[see also,][]{Nordlund1984,Cheung2007}. However, we want to remark that the pattern is less clear than that found in observations \citep[for instance, ][]{Rutten2004,Janssen2006}. At this wavelength, the bright points corresponding to magnetic field concentrations are still brighter than their surroundings. Lastly, if we examine the intensity signals at the line core of both iron lines (rightmost column) we detect a very dynamic landscape  where the granulation pattern is no longer visible and the structures fade away to higher spatial scales. Interestingly, bright photospheric points are still the brightest features and, in some cases, they are surrounded by dark structures that were barely detectable before (see, for instance, the narrow patch at [6,2] Mm). Before moving to the chromospheric lines we would like to mention that we do not detect any appreciable difference between the spatial distribution of intensity signals between the two photospheric lines although the contrast of the features  seems slightly higher for the Fe~{\sc i} 8514~\AA \ line (same colourbar scale).

\begin{figure*}
\centering
 \includegraphics[trim=4 4 3 0,width=17.8cm]{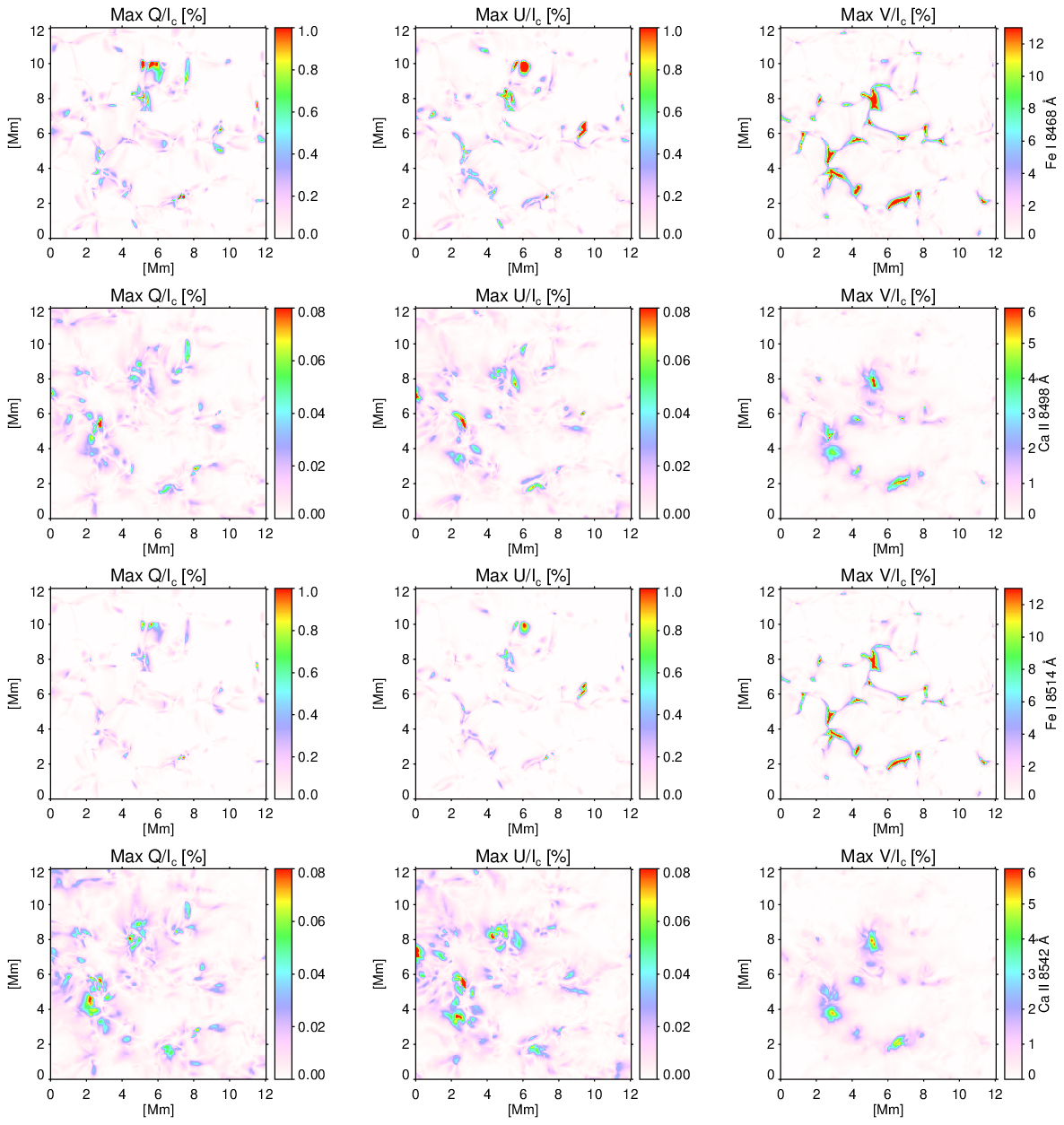}
 \vspace{-0.2cm}
 \caption{Spatial distribution of maximum polarization signals. From left to right, Stokes $Q$, $U$, and $V$ and from top to bottom, the spectral lines with labels 1, 7, 8, and 12 in Table \ref{table_lines}. All values are given as a percentage of the continuum intensity.}
 \label{Polarization}
\end{figure*}

The Ca~{\sc ii} lines are plotted in even rows and we can see that at $\Delta\lambda$=1~\AA, leftmost column, the reverse granulation pattern is slightly visible for the Ca~{\sc ii} 8498~\AA \ line while it is clearer for the broader Ca~{\sc ii} 8542~\AA \ line. It seems that the latter is scanning an atmospheric layer higher than the one examined by the photospheric lines at $\Delta\lambda$=+0.15 \AA \ (middle column) but lower than the one showed by their line core (rightmost column). If we study the intensity signals for the Ca~{\sc ii} lines at $\Delta\lambda$=+0.2 \AA \ we see small differences between them. Both lines show a complex pattern pervaded by elongated features although the contrast  is higher for the Ca~{\sc ii} 8542~\AA \ line. Finally, Ca~{\sc ii} intensity signals are very similar again at their respective line cores. In this case, the landscape shows high contrast structures with bright features that are not always correlated with photospheric magnetic field concentrations and extremely dark areas. This similar landscape means that the Ca~{\sc ii} 8498~\AA \ line core forms at high atmospheric layers, the same layers where the Ca~{\sc ii} 8542~\AA \ forms. Therefore, scanning both lines we will obtain a larger number of spectral points that contain chromospheric information.

\subsection{Maximum polarization signals}

We want to study in the present section whether the polarization Stokes profiles exhibit the same behaviour. To this end, we computed the maximum polarization signals of each Stokes profiles for the four representative lines used in the previous sections. The absolute value of these maximum polarization signals is plotted in Figure \ref{Polarization}, where each column represents a different Stokes profile, from left to right, $Q$, $U$, and $V$, and each row corresponds to a different spectral line organized in the same order as in Figure \ref{Intensity}, i.e. Fe~{\sc i} 8468, Ca~{\sc ii} 8498, Fe~{\sc i} 8514, and Ca~{\sc ii} 8542~\AA, from top to bottom, respectively.

If we first start with the two Fe~{\sc i} photospheric lines, odd rows, we can see that, as happen for the line intensity, the spatial distribution of signals is very similar for both lines. There are linear polarization signals close to the network patch and at [5.6,9.5] Mm, and circular polarization signals in the regions of strong longitudinal magnetic field. However, the larger effective Land\'{e} factor of the Fe~{\sc i} 8468~\AA \ line, produces higher polarization signals for the three Stokes parameters (the colourbar scale of the plots is the same). 

Regarding the Ca~{\sc ii} lines we see a distinct behaviour for each one. The Ca~{\sc ii} 8498 \AA \ line, second row, shows linear polarization signals that are located at the edges of the photospheric longitudinal concentrations but not in the same location as the ones found for the iron lines. They are located outside these photospheric signals indicating that the line is sensitive to the magnetic field at higher layers, where it is more inclined (due to the reduction of the gas pressure the magnetic field expands filling larger areas). Remarkably, the linear polarization signals for the Ca~{\sc ii} 8542~\AA, although similar, show higher amplitude and seem to extend even further from the longitudinal photospheric magnetic field. This is probably a consequence of the different formation heights and Zeeman splitting patterns between the two lines. Moving into the longitudinal fields we found a mixed behaviour for the Ca~{\sc ii} 8498 \AA \ line with regions that closely resemble the photospheric lines, for instance at [6.5,2] Mm, but also spatial locations where the magnetic features occupy a much larger area than those showed by the photospheric lines, e.g. the patch at [3,3.5] Mm. On the contrary, the Ca~{\sc ii} 8542~\AA \ line, always displays a behaviour where the narrow photospheric magnetic field concentrations have turned into large diffuse areas that extend far away from their photospheric roots.

\begin{figure*}
 \includegraphics[trim=2 0 3 0,width=17.5cm]{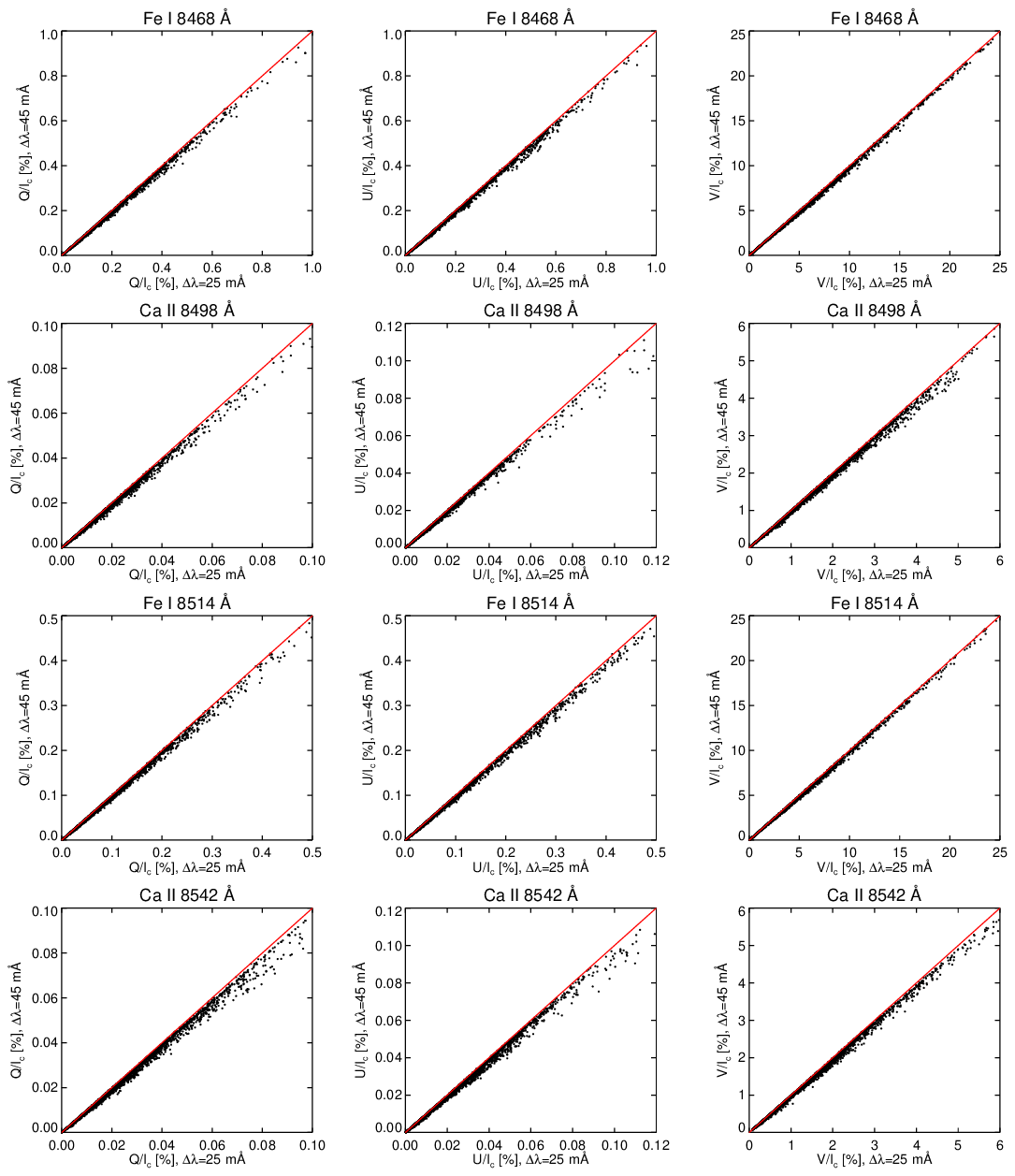}
 \vspace{-0.2cm}
 \caption{Maximum polarization signal using a 25 m\AA \ spectral sampling (x-axis)  versus the case of using a 45 m\AA \ spectral step (y-axis). Red line indicates the case where maximum polarization signals are equal, i.e. $x=y$. Each column designates a different Stokes parameter, being from left to right, Stokes $Q$, $U$, and $V$, while each row corresponds to a different line of Table \ref{table_lines}, with labels 1, 7, 8, and 12.}
 \label{Sampling}
\end{figure*}

\subsection{Spectral sampling effect on the maximum polarization signals}

We demonstrated in Section \ref{1dmodel} that adding all the spectral lines of Table \ref{table_lines} we can infer with higher accuracy the atmospheric parameters at both photospheric and chromospheric layers. However, there is a drawback when we move from observing only one spectral line to measuring a much larger spectral window and that is the spectral sampling of the observation. We used in \cite{QuinteroNoda2016} a spectral window of 8 \AA \ to synthesize the Ca~{\sc ii} 8542 \AA \ line while we propose in the present work a spectral window of 90 \AA, i.e. more than 10 times larger. In principle, we can cover this spectral window with different spectral sampling steps, depending on the number of pixels available on the instrument's camera. For instance, using a $\Delta\lambda=25$ m\AA \ we would need 3600 spectral points, i.e. pixels in the camera, to cover the entire spectrum. However, using a different spectral sampling, e.g. $\Delta\lambda=45$~m\AA \, we would only need 2000 spectral points. This means, that the second case could work with a 2k $\times$ 2k camera while the first one requires at least a 4k $\times$ 4k camera which represents a much higher cost. 

Bearing this limitation in mind we decided to study the impact of increasing the spectral sampling of the observations. There are several ways to do this and we decided to follow a simple approach examining how the amplitude of the polarization signals changes when we used a large spectral sampling (poorer resolution). However, we should note that not only the maximum polarization signals are affected but also the spectral shape of the Stokes profiles could be altered if we use a higher spectral sampling value as we are averaging the signals along the spectral dimension, although this can also help to achieve a lower noise level. 

We synthesized the Stokes profiles for the most relevant lines of Table \ref{table_lines}, i.e. labels 1, 7, 8, and 12. As we mentioned in Section \ref{3dmodel}, we used an ideal spectral sampling of 1 m\AA \ for the synthesis of the Stokes profiles. Then, we degraded the Stokes parameters computing the mean intensity value of each profile along discrete spectral steps. In this case, trying to simplify the visualization of this study, we only used two discrete spectral values whose sizes are 25 and 45 m\AA. In that sense, we aimed to reproduce, in a simplified way, the instrumental effect of degrading the spectral sampling. The input atmospheric parameters are the same as in the previous section and correspond to the {\sc bifrost} simulation snapshot displayed in Figure \ref{fov}. The comparison between maximum polarization signals for a spectral sampling of 25 and 45~m\AA \ are plotted in Figure \ref{Sampling}. We used as a reference the polarization signals obtained with $\Delta\lambda=25$~m\AA \ instead of the original 1~m\AA \ because we consider that, except some spectrographs that have a specific science target that requires extremely high spectral resolution ($\Delta\lambda\sim1$m\AA) as ZIMPOL \citep{Povel2001} or FTS \citep{Brault1979} (only intensity measurements), solar instruments employ a spectral sampling that ranges between $\Delta\lambda=20\sim30$~m\AA \ (see the introduction and Sec. \ref{spectral}).
 
In this occasion, there are not notable differences between the four spectral lines (see Figure \ref{Sampling}). We found that the scatter points followed the red line for lower polarization values for every Stokes parameter. However, as we move to higher polarization signals, the scatter points deviate from the red line towards the x-axis, i.e. $\Delta\lambda=25$~m\AA, indicating that we would detect a lower signal using the poorer spectral sampling of $\Delta\lambda=45$~m\AA. We computed how large could be the difference, in order to account for the worst scenario, and we found a diminishing of up to 17 per cent of the signal obtained using $\Delta\lambda=25$~m\AA. However, the mean loss of signal is smaller, being always between $4\sim6$ per cent for all the lines and Stokes parameters. Thus, the degradation suffered from employing a spectral sampling of $\Delta\lambda=45$~m\AA \ is small probably because the Doppler width of the examined lines is larger than this value.

\section{Summary}

Several studies have demonstrated that the Ca~{\sc ii} 8542~\AA \ line is one of the most promising candidates for performing polarimetric observations, \citep[e.g.,][]{Cauzzi2008}, as it provides continuous information from the photosphere to the chromosphere. In this work, we went a step further and we discovered a design that improves the capabilities of this line adding the neighbouring spectral lines, mainly the second line of the infrared triplet, i.e. Ca~{\sc ii} 8498 \AA \ line, and two strong photospheric lines, i.e. the Fe~{\sc i} 8468 and 8514 \AA \ lines (never used in the literature as far as we know). This new design greatly enhances the sensitivity to the atmospheric parameters at lower heights, thanks to the iron lines, but we also discovered a moderate improvement in the upper atmosphere where the Ca~{\sc ii} 8498 \AA \ line core forms. Moreover, all these lines can be synthesized/inverted simultaneously with {\sc nicole} assuming NLTE and complete redistribution which largely simplifies the computational process. Thus, it is difficult to find a strong argument against observing all the lines of Table~\ref{table_lines} to improve the sensitivity to the atmospheric parameters. The only one is the increase of the spectral sampling (poorer spectral resolution) of the observation. However, we performed a quick test finding small differences on the maximum polarization signals between using a reference spectral sampling of $\Delta\lambda=25$~m\AA \ or a worse value of $\Delta\lambda=45$~m\AA. Therefore, we can conclude that one of the best options to perform chromospheric polarimetry is to measure all the lines included in Table \ref{table_lines} and, for this reason, we encourage future missions as Solar-C, DKIST, and EST to observe them.

As future work, we plan to perform an additional study to confirm the advantages of measuring multiple lines. We aim to compare the results from the inversion of the Stokes profiles generated from an artificial atmospheric toy model when we use only the Ca~{\sc ii} 8542 \AA \ line or we invert all the lines of Table \ref{table_lines}. We foresee an improvement in the atmospheric parameters at lower heights based on the RF results presented in Figure \ref{1DRF}. However, we also want to quantify the expected improvement at higher layers as we will increase the number of wavelength points that are sensitive to these heights. In addition, it is well known that the isotopic splitting of the Ca~{\sc ii} 8542 \AA \ line affects the retrieved gradients of the atmospheric parameters \citep{Leenaarts2014}, thus we aim to perform a similar study as the one done by these authors for the  Ca~{\sc ii} 8498 \AA \ line to establish how important is the isotopic splitting in this case. Moreover, it will be interesting to check the differences between 1D and 3D computations following the steps of \cite{Sukhorukov2016} although in our case it will also be necessary to estimate this effect on the polarization profiles. Finally, we also plan to perform observations of different solar features, for instance, a sunspot or a plage region, to estimate the polarization signals produced by all the lines of the 850 nm spectral region.

\section*{Acknowledgements}
We thank B. Ruiz Cobo and H. Socas Navarro for their valuable comments. We also thank H. Uitenbroek because he was the one that proposed us to examine the other Ca~{\sc ii} lines of the infrared triplet to improve our capabilities for chromospheric polarimetry. This work was supported by MEXT/JSPS KAKENHI Grant Numbers JP15H05814 and JP15K21709. JdlCR is supported by grants from the Swedish Research Council (VR) and the Swedish National Space Board (SNSB). The research leading to these results has received funding from the European Research Council under the European Union's Seventh Framework Programme (FP7/2007-2013)/ERC grant agreement no 291058.

\bibliographystyle{mnras} 
\bibliography{multiline} 

\bsp	
\label{lastpage}
\end{document}